# Self-doping effect in confined copper selenide semiconducting quantum dots for efficient photoelectrocatalytic oxygen evolution

Ce Hu[1,3,#], Jie Ren[1,2,#], Daojian Ye[1,2,#], Chenya Zhao[1,2], Lanshan He[1,2], Congcong Wu[1,2], Wenting Jia[1,2], Shengwen Xu[1,2], Weiyang Xu[1,2], Fujin Huang[1,2], Hang Zhou[1,2], Chengwu Zou[1,2], Ting Yu[1,2], Xingfang Luo[1,2] and Cailei Yuan[1,2,*]

[1]Jiangxi Key Laboratory of Nanomaterials and Sensors, Jiangxi Normal University, 99 Ziyang Avenue, Nanchang 330022, Jiangxi, China.

[2]School of Physics, Communication and Electronics, Jiangxi Normal University, 99 Ziyang Avenue, Nanchang 330022, Jiangxi, China.

[3]Analytical & Testing Center, Jiangxi Normal University, 99 Ziyang Avenue, Nanchang 330022, Jiangxi, China.

*Corresponding author. E-mail: clyuan@jxnu.edu.cn

[#]C.H., J.R. and D.Y. contributed equally to this work.

# Abstract

Self-doping (i.e. in situ defect engineering) can not only suppress the photogenerated charge recombination of semiconducting quantum dots by self-introducing trapping states within the bandgap, but also provide high-density catalytic active sites as the consequence of abundant non-saturated bonds associated with the defects. And the construction of confined structure has emerged as an attractive strategy to accomplishing efficient charge transfer in dispersed semiconducting quantum dots, which can not only realize ultra-high conductivity by fully contacting with the conductive confined host, but also can protect quantum dots from degradation to obtain ultra-high stability of photoelectrocatalysis. Based on the forementioned two sides, here, we successfully prepared semiconducting copper selenide (CuSe) confined quantum dots with abundant vacancies and systematically investigated their photoelectrochemical characteristics. Photoluminescence characterizations reveal that the presence of vacancies reduces the emission intensity dramatically, indicating a low recombination rate of photogenerated charge carriers due to the self-introduced trapping states within the bandgap. In addition, the ultra-low charge transfer resistance measured by electrochemical impedance spectroscopy implies the efficient charge transfer of CuSe semiconducting quantum dots-based photoelectrocatalysts, which is guaranteed by the high conductivity of their confined structure as revealed by room-temperature electrical transport measurements. Such high conductivity and low photogenerated charge carriers recombination rate, combined with high-density active sites and confined structure, guaranteeing the remarkable photoelectrocatalytic performance and stability as manifested by photoelectrocatalysis characterizations. This work promotes the development of semiconducting quantum dots-based photoelectrocatalysis and demonstrates CuSe semiconducting quantum confined catalysts as an advanced photoelectrocatalysts for oxygen evolution reaction.

**Keywords:** Self-doping effect; confined structure; semiconducting quantum dots; copper selenide; photoelectrocatalytic oxygen evolution.

**Introduction**

Photoelectrocatalysis, facilitating the catalytic reaction by photogenerated electrons and holes under natural sunlight illumination, is a clean and sustainable technique for various electrochemical energy storage and conversion technologies as water splitting, $CO_2$ reduction, and pollutant degradation.[1-3] Among various semiconductor-based photoelectrocatalysts, semiconducting quantum dots have raised significant attention due to their quantum size effect, large specific surface area and high quantum yield.[4-6] Unfortunately, the high recombination rate of photogenerated charge carriers severely limits the photoelectrocatalytic performance of semiconducting quantum dots.[7-9] In this regard, self-doping (i.e. in situ defect engineering) is expected to break this limitation.[10-13] In situ engineering defects in semiconducting quantum dots can not only suppress the photogenerated charge recombination by self-introducing trapping states within the bandgap, but also provide high-density catalytic active sites as the consequence of abundant non-saturated bonds associated with the defects.[14-21]

On the other side, the mono-dispersity of semiconducting quantum dots naturally determines their poor conductivity, and thus hindering catalytic kinetics fundamentally.[22,23] For the purpose of realizing efficient charge transfer in semiconducting quantum dots, an effective method is to disperse quantum dots on conductive supporting framework.[24-28] Under this research background, the confined structure can not only provide researchers with an effective method to achieve good uniformity, large density and especially high conductivity of semiconducting quantum dots by contacting with the conductive confined host, but also can protect the semiconducting quantum dots from degradation to obtain ultra-high stability of photoelectrocatalysis.[29-31] Such confined semiconducting quantum dots will provide an excellent platform for achieving efficient and robust photoelectrocatalysis under solar illumination.

Here, in terms of the two forementioned sides of self-doping effect and confined structure, we successfully synthesized copper selenide (CuSe) confined semiconducting quantum dots with abundant selenium vacancies and systematically investigated their photoelectrochemical characteristics. Photoluminescence (PL) characterizations show that the presence of vacancies dramatically reduces the emission intensity, indicating a low recombination probability of photogenerated charge carriers owing to the self-introduced trapping states within the bandgap (**Figure 1**a). Besides, an ultra-low charge transfer resistance measured by electrochemical impedance spectroscopy (EIS) suggests the efficient charge transfer of CuSe semiconducting quantum dots-based photoelectrocatalysts, which is guaranteed by the high conductivity of their confined structure as confirmed by room-temperature electrical transport measurements (Figure 1b). Thanks to the low electron-hole pair recombination probability and high electrical conductivity, together with the high-density catalytic active sites and confined structure, the prepared defective CuSe semiconducting quantum dots exhibit superior photoelectrocatalytic oxygen evolution reaction (OER) performance and excellent stability as illustrated by photoelectrocatalysis characterizations. This work proposed an effective confined framework to achieve efficient and robust photoelectrocatalysts based on dispersed semiconducting quantum dots and promoted the development of semiconducting quantum dots-based photoelectrocatalysis by self-doping effect.

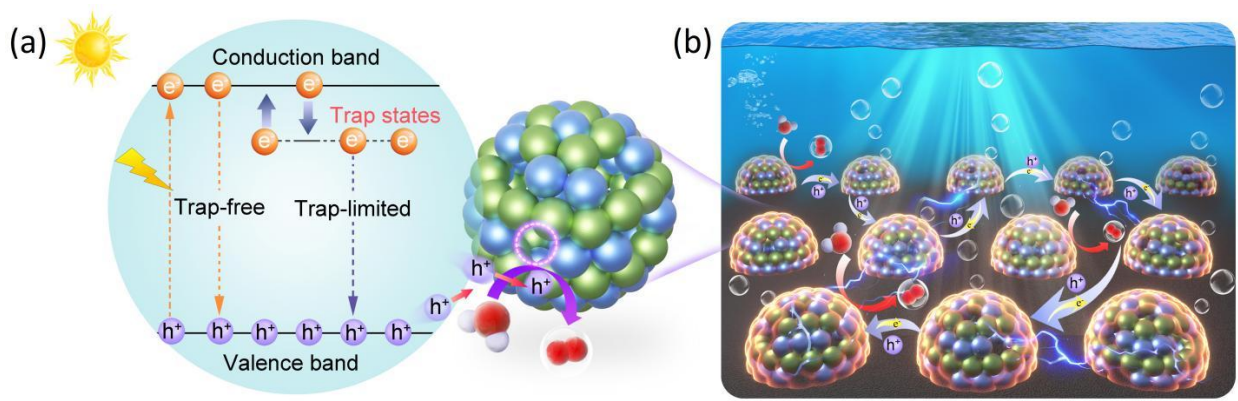

**Figure 1.** Schematic diagram of **(a)** the suppression of photogenerated charge recombination and **(b)** efficient charge transfer in dispersed semiconducting quantum dots.

## Results and discussion

The self-doped CuSe quantum confined catalysts i.e. semiconducting CuSe quantum dots/conductive amorphous carbon matrix nanocomposites were prepared via pulsed laser deposition (PLD) technique (detailed information of catalyst preparation and characterizations are given in the Experimental Methods of **Supplementary information**). Due to high substrate temperature and ultrahigh vacuum conditions of the PLD deposition, abundant selenium vacancies were in situ generated in CuSe semiconducting quantum dots, and resulting in the self-doping effect. The data displayed in this paper were acquired from the representative self-doped CuSe quantum confined catalysts with optimal photoelectrocatalytic performance, which possesses the smallest quantum dot size, by virtues of high specific surface area and strong self-doping effect guaranteed good electrical conductivity as well as better photoelectrochemical efficiency.

Prior to investigating the photoelectrochemical performance of PLD-prepared self-doped confined catalysts, we systematically characterized the nanostructural and photoelectronic characteristic of the selenium vacancies-abundant CuSe semiconducting quantum dots. We start with characterizing the structural properties by transmission electron microscopy (TEM). As shown in **Figure 2**a, a representative low-resolution TEM (LRTEM) image, illustrated that the preparation of high-density and size-homogeneous CuSe quantum dots confined in amorphous carbon matrix have been achieved. Figure S1 shows the LRTEM images at other magnifications. The well-dispersed CuSe quantum dots present nearly spherical structures with high crystalline quality. After statistical analysis of the LRTEM results, it can be obtained that the areal density of CuSe quantum dots is about $1.9 \times 10^{12}$ cm$^{-2}$, which is close to the ultimate density of compact packing. And the average diameter of CuSe quantum dots is around 4.15 nm. Besides, the TEM equipped selected region electron diffraction (SAED) was also utilized to examine the crystal structure and crystalline quality. As shown in Figure 2b, the representative SAED pattern is well consistent with the diffraction pattern of the hexagonal space group *P63/mmc* simulated by Java electron microscopy simulation (JEMS) software, indicating that the PLD-prepared CuSe quantum dots have hexagonal crystal structure. Furthermore, as shown in Figure 2c, the typical high-resolution TEM (HRTEM) image of a well-defined single quantum dot demonstrated the single-crystal feature with a 0.32 nm interplanar spacing, which can be identified as the CuSe (102) crystallographic plane.[32] Moreover, the structural property and phase composition of the CuSe quantum dots were further investigated by X-ray diffraction (XRD) characterizations.

Figure 2d shows a representative XRD pattern of the CuSe quantum confined catalysts. All the diffraction peaks can be well indexed to the hexagonal phase CuSe reported in JCPDS card (No. 34-0171).[33] And no diffraction peaks associated with impurities or heterophase were found, confirming the pure phase of confined CuSe quantum dots and the amorphous nature of carbon matrix. Additionally, the strong diffraction peak assigned to the (102) plane implies that PLD-prepared CuSe quantum dots have a highly preferential crystal orientation. The pure hexagonal phase with klockmannite structure and the preferential crystallite orientation of (102) plane concluding from the XRD characterization is in good agreement with the TEM and SAED results.

For photoelectrocatalysis, an appropriate optical band gap of semiconducting catalyst is vital for a good photoresponse to solar light. Thus, the ultraviolet-visible (UV-vis) absorption spectroscopy was utilized to examine the energy band gap of PLD-prepared CuSe semiconducting quantum dots. Figure 2e and its inset respectively shows the full-range UV-vis absorption spectrum and the corresponding Tauc-Mott plot. As shown in Figure 2e, the CuSe quantum dots exhibit wide-spectrum visible light absorption. By extrapolating the linear portion of the Tauc-Mott plot to intersect the photon energy axis, it can be determined that the band gap of PLD-prepared semiconducting quantum dots is 2.50 eV, which correspond to the direct band gap of CuSe usually reported in the literature.[34-36] Therefore, good solar-energy harvesting property of the CuSe semiconducting quantum dots can be expected from the visible range direct band gap.

In catalytic reactions, structural defect sites in materials are generally considered to be the key active sites. And defect engineering has emerged as a key strategy to prepare advanced catalysts. Benefiting from the high substrate temperature and ultrahigh vacuum conditions of PLD technique, abundant selenium vacancies were in situ formed at CuSe quantum dots synthesis stage. To determinate the chemical stoichiometry and valence state of PLD-prepared CuSe semiconducting quantum dots, X-ray photoelectron spectroscopy (XPS) characterizations were conducted. Figure S2 displays the overall XPS spectrum. As shown in Figure 2f (Cu 2p region), the doublet peaks at 952.5 and 932.6 eV are associated with Cu $2p_{1/2}$ and Cu $2p_{3/2}$ respectively, indicating the presence of Cu(I).[37] And the another doublet peaks for Cu $2p_{1/2}$ and Cu $2p_{3/2}$ locate at 954.0 and 933.5 eV along with the satellite lines can be assigned to Cu(II). In the Se 3d region (Figure 2g), the doublet peaks at at 55.1 and 54.2 eV can be attributed to Se $3d_{3/2}$ and Se $3d_{5/2}$ respectively. Based on the XPS characterizations, the atomic concentration ratio of Cu to Se is calculated to be 1.37:1, suggesting the PLD-prepared CuSe semiconducting quantum dots are selenium vacancies-abundant. Consequently, we can conclude that high density semiconducting CuSe confined quantum dots with abundant selenium vacancies have been successfully prepared.

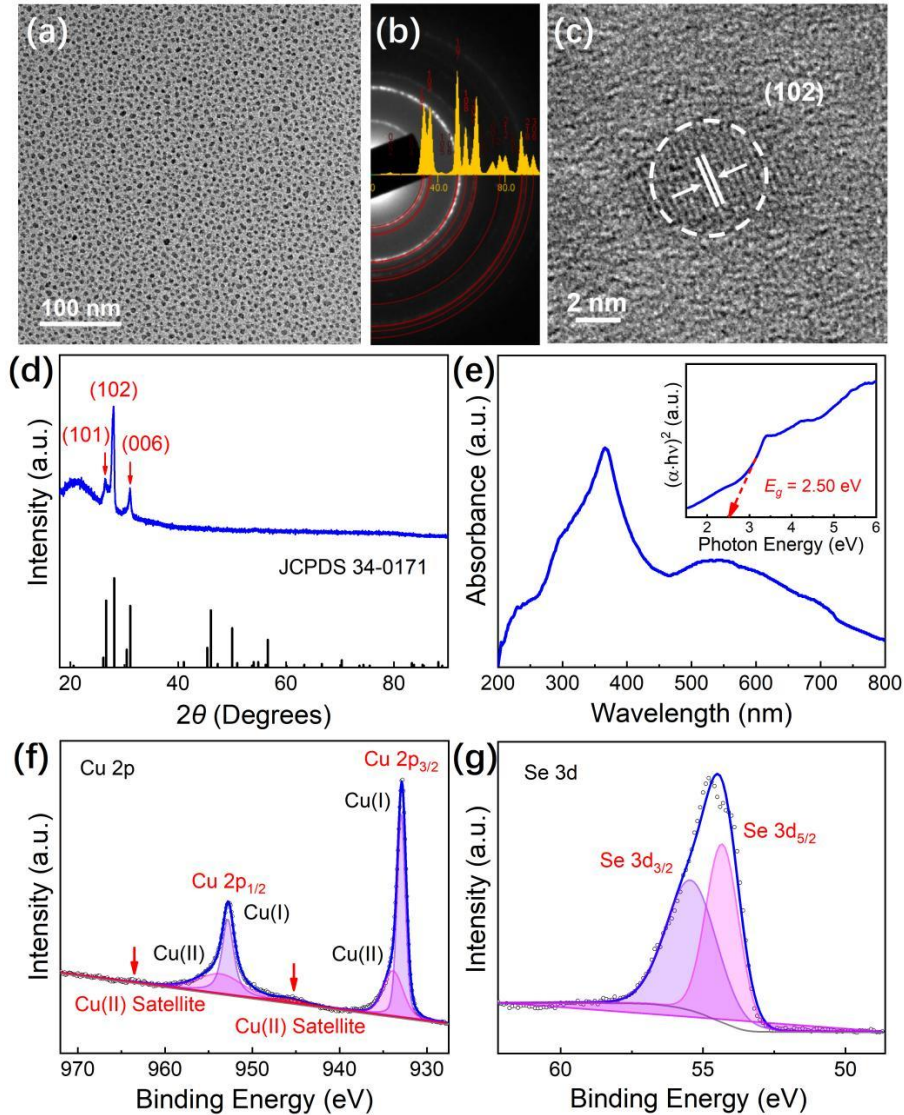

**Figure 2. (a)** LRTEM image and **(b)** the corresponding SAED pattern of CuSe quantum dots. **(c)** HRTEM image of a representative single quantum dot. **(d)** XRD pattern, **(e)** UV-vis absorption spectrum (inset, the corresponding Tauc-Mott plot) and **(f, g)** XPS spectra of self-doped CuSe quantum confined catalysts.

The presence of selenium vacancies in PLD-prepared CuSe quantum dots was further confirmed by electron paramagnetic resonance (EPR) characterizations. As shown in **Figure 3**a, the EPR spectrum shows an obvious signal at $g$ = 2.015, corresponding to the characteristic peak of selenium vacancies.[38] The abundant selenium vacancies in CuSe semiconducting quantum dots will lead to the self-doping effect and further alter its optoelectronic characteristics significantly. In order to investigate the influence of self-doping effect on the properties of CuSe quantum dots, we obtained pristine CuSe quantum dots as contrast sample by annealing in selenium atmosphere to heal the vacancies (detailed information are given in **Experimental Methods**). As shown in Figure 3a, the EPR signal at $g$ = 2.015 is notably suppressed, implying that most selenium vacancies have been filled. Since the valence band structure near the Fermi level is crucial to the OER kinetics, we collected the valence band spectra of CuSe quantum dots.[39] The valence band maximum (VBM) values are determined by extrapolating the linear portion of the valence band spectra to intersect the binding energy axis. As shown in Figure 3b, with the presence of abundant selenium vacancies in CuSe semiconducting quantum dots, VBM value increases from 1.76 to 2.01 eV as the consequence of n-type self-doping. The abundant selenium vacancies, acting as electron donors, will lead to the Fermi level shifts toward the conductive band minimum, and the electrical conductivity is thus increased correspondingly which is experimentally confirmed by the room temperature electrical transport measurements. According to the electrical transport characterizations (Figure S3), an ultra-low resistivity of 2.32 μΩ·m is determined, revealing the high electrical conductivity of selenium vacancies-abundant CuSe semiconducting quantum dots (i.e., self-doped CuSe quantum confined catalysts).

For advanced photoelectrocatalysts, in addition to the high conductivity feature, the low recombination probability of photogenerated charge carriers is also important. Figure 3c shows the room-temperature PL spectra of the pristine and self-doped CuSe semiconducting quantum dots. All the PL spectra display wide-spectrum visible light luminescence. However, with the presence of abundant selenium vacancies (viz, self-doping effect), the PL intensity decreases significantly. It is well-known that PL emission of a semiconductor is the consequence of recombination of photogenerated electron-hole pairs.[40] Thus, the lower PL intensity of self-doped CuSe semiconducting quantum dots indicates a lesser probability of electron-hole pairs' recombination due to the self-introduced trapping states within the bandgap. It is conceivable that such high-conductive CuSe quantum confined catalysts with low recombination probability of photogenerated electron-hole pairs can exhibit a good photoelectrocatalytic performance for oxygen evolution.

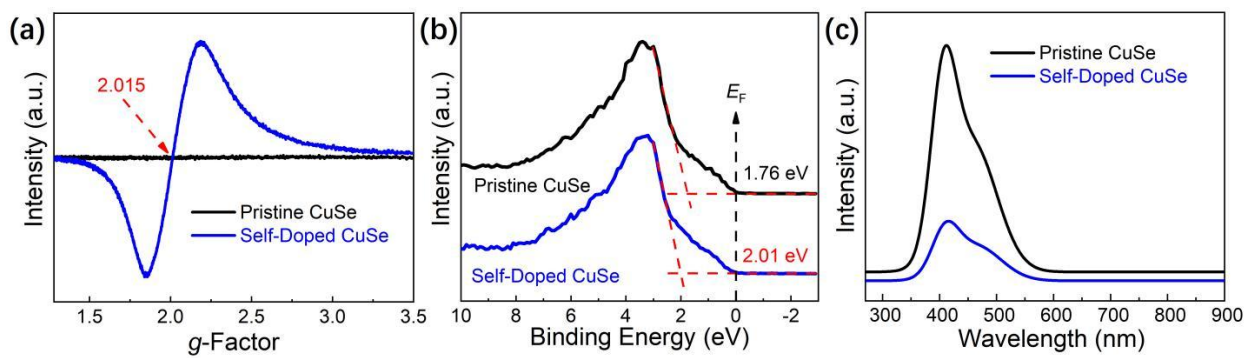

**Figure 3. (a)** EPR, **(b)** valence band and **(c)** PL spectra of pristine and self-doped CuSe quantum confined catalysts.

Then the photoelectrocatalytic performance of self-doped CuSe quantum confined catalysts was studied by electrochemical measurements. The commercial $IrO_2$ electrode used as a reference catalyst was measured, and the OER performance (**Figure 4**a) was similar to that reported in the literature, indicating the reliability of our electrochemical measurement data. As shown in Figure 4a and 4b, even without the solar illumination, the electrocatalytic performance of self-doped CuSe quantum dots for OER is comparable to that of current state-of-the-art CuSe catalysts, which can drive a current density of 100 mA/cm$^2$ at a low overpotential of −434 mV with a low Tafel slope of 70.0 mV/dec.[40,41] More importantly, due to the superior photoelectronic properties of self-doped CuSe semiconducting quantum dots, the OER performance can be further improved with solar illumination. Under solar illumination, the self-doped CuSe quantum confined catalysts displayed a smaller overpotential of −403 mV and a lower Tafel slope of 64.1 mV/dec, indicating the accelerated OER kinetic process and good photoelectrochemical efficiency. In addition, the electron transfer process occurring at the electrode-solution interface of the self-doped CuSe quantum confined catalysts was investigated using EIS measurements. Interpretation of EIS measurements is usually done by correlation between impedance data and equivalent circuits, which represent the physical processes occurring in the system. Randles equivalent circuits are widely used in many electrochemical systems. In view of the non-ideal characteristics of capacitor, a constant phase element (CPE) is introduced into the equivalent circuit to improve the measurement accuracy of capacitance and resistance. The equivalent Randles circuit model (Figure S4) was used to fit the Nyquist diagram to determine the charge transfer resistance ($R_{ct}$) and solution impedance ($R_s$). By simulating the impedance data (Figure 4c), the obtained $R_{ct}$ with solar illumination (about 9.3 Ω) is lower than that of without illumination (about 14.5 Ω), indicating a higher electron transfer efficiency and the boosted mass transport. Furthermore, time-resolved response measurement was utilized to investigate the photo-switching behavior of self-doped CuSe quantum confined catalysts. As shown in Figure 4d, the OER current density has a reversible response to the periodic solar illumination, revealing reproducible optical switching behavior. As a result, the largely improved catalytic performance toward OER induced by solar illumination, has been demonstrated in the self-doped CuSe quantum confined catalysts.

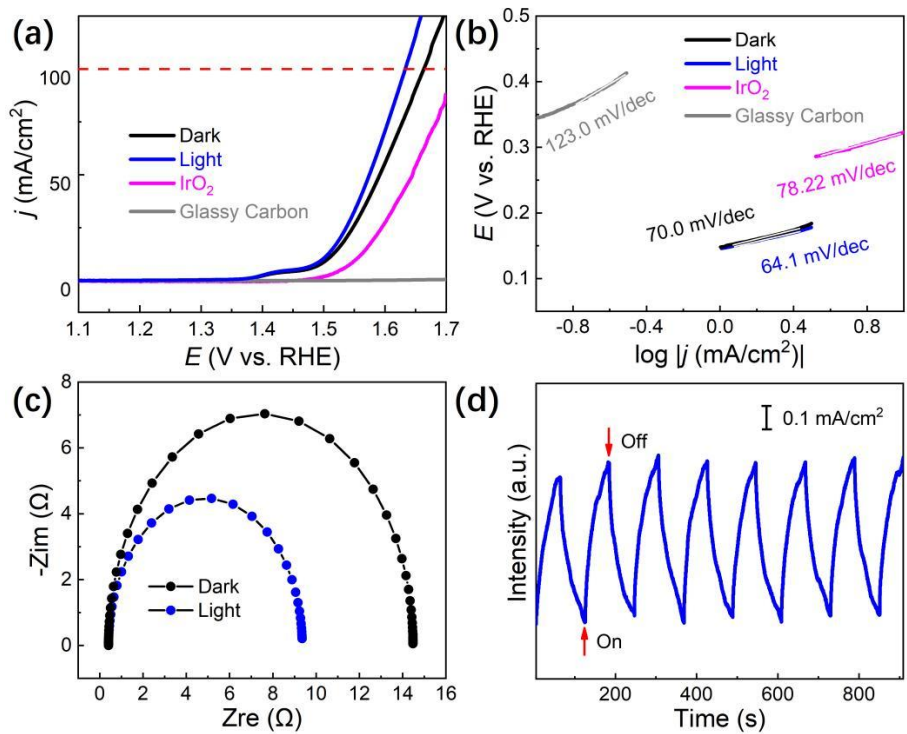

**Figure 4. (a)** LSV curves, **(b)** corresponding Tafel plots and **(c)** Nyquist plots of self-doped CuSe quantum confined catalyst with and without solar illumination. **(d)** Time-resolved photo-response to the periodic solar illumination.

More importantly, besides to the high photoelectrocatalytic activity, as shown in **Figure 5**a, the self-doped CuSe quantum confined catalysts showed consistent OER characteristics under continuous operations of 10,000 cycles, which indicates its high catalytic stability due to the confined structure. In addition, we performed Raman and XPS characterizations to further evaluate the structural stability of CuSe quantum dots. As shown in Figure 5b, the representative Raman spectra of self-doped CuSe quantum dots before and after long-term OER operation show three identical characteristic modes of CuSe, located at 196.0, 261.2 and 517.2 cm$^{-1}$, which are corresponding to LO (first longitudinal optic), TO (transverse optic) and 2LO (second longitudinal optic) phonon modes respectively.[42] Moreover, compared with the XPS spectrum of CuSe quantum dots before long-term OER operation, the spectral characteristics (peak position, half-width and peak intensity) did not change after stability test (Figure 5c and 5d), indicating that the changes in chemical composition and valence states were negligible. Therefore, the high activity durability and structural stability of PLD-prepared self-doped CuSe quantum confined catalysts were confirmed.

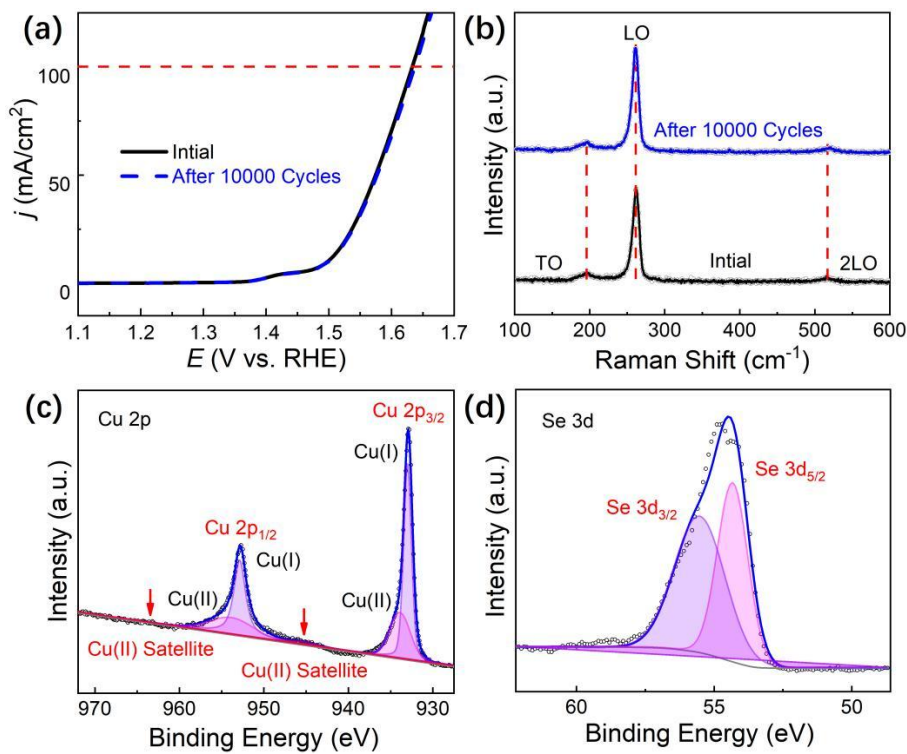

**Figure 5. (a)** LSV curves and **(b)** Raman spectra before and after 10,000 cycles of continuous OER operations. XPS spectra after long-term operation, showing **(c)** Cu 2p and **(d)** Se 3d core levels.

## Conclusions

In summary, based on the preparation of selenium vacancies-abundant CuSe semiconducting confined quantum dots, the influences of self-doping effect and confined structure on the photoelectrocatalytic characteristics are investigated systematically. The PL characterizations reveal that the presence of abundant selenium vacancies dramatically reduces the emission intensity, indicating a low probability of electron-hole pairs' recombination owing to the self-introduced trapping states within the bandgap. Besides, an ultra-low charge transfer resistance (about 9.3 Ω) measured by EIS suggests the efficient charge transfer of CuSe semiconducting quantum dots-based photoelectrocatalysts, which is guaranteed by the high conductivity (low resistivity of 2.32 μΩ·m) of their confined structure as confirmed by room-temperature electrical transport measurements. The high electrical conductivity and low photo-excited charge carriers recombination rate, combined with high density active sites and confined structure, guaranteeing the remarkable and stable photoelectrocatalytic performance of the self-doped CuSe semiconducting quantum dots as manifested by photoelectrocatalysis characterizations. Under solar illumination, displaying a small overpotential of −268 mV to drive a current density of −10 mA/cm$^2$ with a low Tafel slope of 64.1 mV/dec. This work advanced the development of semiconducting quantum dots-based photoelectrocatalysis by self-doping effect and proposes an effective confined framework to achieve efficient and robust photoelectrocatalysts based on dispersed semiconducting quantum dots.

## Competing interests

## Acknowledgements

This work was supported by the National Natural Science Foundation of China (Grant No. 12204203, 52061017, 52171213, 51871115 and 62265009), Project of Academic and Technological Leaders in Jiangxi Province (Grant No. 20213BCJ22010 and 20204BCJ22036), Natural Science Foundation of Jiangxi Province (Grant No. 2021BAB211014), and Research Projects of Education Department of Jiangxi Province (Grant No. 191700).

## Data availability

The data that support the findings of this study are available from the corresponding author upon reasonable request.

## Supplementary information

Supplementary information is available for this paper, including Experimental methods, LRTEM images, overall XPS spectrum, *I-V* curve at room temperature, an equivalent circuit to fit the EIS of OER process and calibration of reference electrode.

## References

1. Yao, T., An, X., Han, H., Chen, J. Q. & Li, C. Photoelectrocatalytic materials for solar water splitting. *Adv. Energy Mater.* **8**, 1800210 (2018).

2. Shan, B., Vanka, S., Li, T., Troian-Gautier, L., Brennaman, M. K., Mi, Z. & Meyer, T. J. Binary molecular-semiconductor p-n junctions for photoelectrocatalytic $CO_2$ reduction. *Nat. Energy* **4**, 290-299 (2019).

3. Song, R., Chi, H., Ma, Q., Li, D., Wang, X., Gao, W., Wang, H., Wang, X., Li, Z. & Li, C. Highly efficient degradation of persistent pollutants with 3D nanocone $TiO_2$-based photoelectrocatalysis. *J. Am. Chem. Soc.* **143**, 13664-13674 (2021).

4. Xu, Z. L., Lin, S., Onofrio, N., Zhou, L., Shi, F., Lu, W., Kang, K., Zhang, Q. & Lau, S. P. Exceptional catalytic effects of black phosphorus quantum dots in shuttling-free lithium sulfur batteries. *Nat. Commun.* **9**, 1-11 (2018).

5. Wang, Y., Ma, Y., Li, X. B., Gao, L., Gao, X. Y., Wei, X. Z., Zhang, L. P., Tung, C. H., Qiao, L. & Wu, L. Z. Unveiling catalytic sites in a typical hydrogen photogeneration system consisting of semiconductor quantum dots and 3d-metal ions. *J. Am. Chem. Soc.* **142**, 4680-4689 (2020).

6. Amirav, L. & Alivisatos, A. P. Luminescence studies of individual quantum dot photocatalysts. *J. Am. Chem. Soc.* **135**, 13049-13053 (2013).

7. Wang, X., Sun, G., Li, N. & Chen, P. Quantum dots derived from two-dimensional materials and their applications for catalysis and energy. *Chem. Soc. Rev.* **45**, 2239-2262 (2016).

8. Caputo, J. A., Frenette, L. C., Zhao, N., Sowers, K. L., Krauss, T. D. & Weix, D. J. General and efficient C-C bond forming photoredox catalysis with semiconductor quantum dots. *J. Am. Chem. Soc.* **139**, 4250-4253 (2017).

9. Moon, B. J., Kim, S. J., Lee, A., Oh, Y., Lee, S. K., Lee, S. H., Kim. T. W., Hong, B. H. & Bae, S. Structure-controllable growth of nitrogenated graphene quantum dots via solvent catalysis for selective CN bond activation. *Nat. Commun.* **12**, 1-11 (2021).

10. Huang, H., Li, X., Wang, J., Dong, F., Chu, P. K., Zhang, T. & Zhang, Y. Anionic group self-doping as a promising strategy: band-gap engineering and multi-functional applications of high-performance $CO_3^{2-}$-Doped $Bi_2O_2CO_3$. *ACS Catal.* **5**, 4094-4103 (2015).

11. Cho, S. H., Byeon, J., Jeong, K., Hwang, J., Lee, H., Jang, J., Lee, J., Kim, T., Kim, K., Choi, M. & Lee, Y. S. Investigation of defect-tolerant perovskite solar cells with long-term stability via controlling the self-doping effect. *Adv. Energy Mater.* **11**, 2100555 (2021).

12. Justicia, I., Ordejón, P., Canto, G., Mozos, J. L., Fraxedas, J., Battiston, G. A., Gerbasi, R. & Figueras, A. Designed self-doped titanium oxide thin films for efficient visible-light photocatalysis. *Adv. Mater.* 14, 1399-1402 (2002).

13. Mao, C., Zuo, F., Hou, Y., Bu, X. & Feng, P. In situ preparation of a $Ti^{3+}$ self-doped $TiO_2$ film with enhanced activity as photoanode by $N_2H_4$ reduction. *Angew. Chem., Int. Ed.* 126, 10653-10657 (2014).

14. Yang, R., Peng, S., Lan, B., Sun, M., Zhou, Z., Sun, C., Gao, Z., Xing, G. & Yu, L. Oxygen defect engineering of β-$MnO_2$ catalysts via phase transformation for selective catalytic reduction of NO. *Small* **17**, 2102408 (2021).

15. Ortiz-Medina, J., Wang, Z., Cruz-Silva, R., Morelos-Gomez, A., Wang, F., Yao, X., Terrones, M. & Endo, M. Defect engineering and surface functionalization of nanocarbons for metal-free catalysis. *Adv. Mater.* **31**, 1805717 (2019).

16. Cao, Y., Sheng, Y., Edmonds, K. W., Ji, Y., Zheng, H. & Wang, K. Deterministic magnetization switching using lateral spin-orbit torque. *Adv. Mater.* **32**, 1907929 (2020).

17. Arandiyan, H., Mofarah, S. S., Sorrell, C. C., Doustkhah, E., Sajjadi, B., Hao, D., Wang, Y., Sun, H., Ni, B., Rezaei, M., Shao, Z. & Maschmeyer, T. Defect engineering of oxide perovskites for catalysis and energy storage: synthesis of chemistry and materials science. *Chem. Soc. Rev.* **50**, 10116-10211 (2021).

18. Cai, K., Yang, M., Ju, H., Wang, S., Ji, Y., Li, B., Edmonds, K. W., Sheng, Y., Zhang, B., Zhang, N., Liu, S., Zheng, H. & Wang, K. Electric field control of deterministic current-induced magnetization switching in a hybrid ferromagnetic/ferroelectric structure. *Nat. mater.* **16**, 712-716 (2017).

19. He, J., Li, N., Li, Z., Zhong, M., Fu, Z., Liu, M., Yin, J., Shen, Z., Li, W., Zhang, J., Chang,

Z. & Bu, X. Strategic defect engineering of metal-organic frameworks for optimizing the fabrication of single-atom catalysts. *Adv. Funct. Mater.* **31**, 2103597 (2021).

20. Zhang, Xi. & Zhang, L. Electronic and band structure tuning of ternary semiconductor photocatalysts by self Doping: the case of BiOI. *J. Phys. Chem. C* **114**, 18198-18206 (2010).

21. Wei, F., Liu, Y., Zhao, H., Ren, X., Liu, J., Hasan, T., Chen, L., Li, Y. & Suacd, B. Oxygen self-doped g-$C_3N_4$ with tunable electronic band structure for unprecedentedly enhanced photocatalytic performance. *Nanoscale* **10**, 4515-4522 (2018).

22. Yeh, T. F., Teng, C. Y., Chen, S. J. & Teng, H. Nitrogen-doped graphene oxide quantum dots as photocatalysts for overall water-splitting under visible light illumination. *Adv. Mater.* **26**, 3297-3303 (2014).

23. Wang, X., Feng, Y., Dong, P. & Huang, J. A mini review on carbon quantum dots: preparation, properties, and electrocatalytic application. *Front. Chem.* **7**, 671 (2019).

24. Kaur, M., Ubhi, M. K., Grewal, J. K. & Sharma, V. K. Boron- and phosphorous-doped graphene nanosheets and quantum dots as sensors and catalysts in environmental applications: a review. *Environ. Chem. Lett.* **19**, 4375-4392 (2021).

25. Yang, H., Yin, J., Cao, R., Sun, P., Zhang, S. & Xu, X. Constructing highly dispersed 0D $Co_3S_4$ quantum dots/2D g-$C_3N_4$ nanosheets nanocomposites for excellent photocatalytic performance. *Sci. Bull.* **64**, 1510-1517 (2019).

26. Wang, Z., Zheng, Z., Xue, Y., He, F. & Li, Y. Acidic water oxidation on quantum dots of $IrO_x$/graphdiyne. *Adv. Energy Mater.* **11**, 2101138 (2021).

27. Gao, F., Zhao, Y., Zhang, L., Wang, B., Wang, Y., Huang, X., Wang, K., Feng, W. & Liu, P. Well dispersed MoC quantum dots in ultrathin carbon films as efficient co-catalysts for photocatalytic $H_2$ evolution. *J. Mater. Chem. A* **6**, 18979-18986 (2018).

28. Xi, J., Xia, H., Ning, X., Zhang, Z., Liu, J., Mu, Z., Zhang, S., Du, P. & Lu, X. Carbon-intercalated 0D/2D hybrid of hematite quantum dots/graphitic carbon nitride nanosheets as superior catalyst for advanced oxidation. *Small* **15**, 1902744 (2019).

29. Shifa, T. A. & Vomiero, A. Confined catalysis: progress and prospects in energy conversion. *Adv. Energy Mater.* **9**, 1902307 (2019).

30. Li, Z., Zhang, X., Cheng, H., Liu, J., Shao, M., Wei, M., Evans, D. G., Zhang, H. & Duan, X. Confined synthesis of 2D nanostructured materials toward electrocatalysis. *Adv. Energy Mater.* **10**, 1900486 (2020).

31. Zhou, Y., Xie, Z., Jiang, J., Wang, J., Song, X., He, Q., Ding, W. & Wei, Z. Lattice-confined Ru clusters with high CO tolerance and activity for the hydrogen oxidation reaction. *Nat. Catal.* **3**, 454-462 (2020).

32. Shi, Z., Sun, Z., Yang, X., Lu, C., Li, S., Yu, X., Ding, Y., Huang, T. & Sun, J. Synergizing conformal lithiophilic granule and dealloyed porous skeleton toward pragmatic Li metal anodes. *Small Science* **2**, 2100110 (2022).

33. Wu, X. J., Huang, X., Liu, J., Li, H., Yang, J., Li, B., Huang, W. & Zhang, H. Two-dimensional CuSe nanosheets with microscale lateral size: Synthesis and template-assisted phase transformation. *Angew. Chem., Int. Ed.* **53**, 5083-5087 (2014).

34. Li, H., Zhu, Y., Avivi, S., Palchik, O., Xiong, J., Koltypin, Y., Palchik, V. & Gedanken, A. Sonochemical process for the preparation of α-CuSe nanocrystals. *J. Mater. Chem.* **12**, 3723-3727 (2002).

35. Al-Mamun & Islam, A.B.M.O. Characterization of copper selenide thin films deposited by chemical bath deposition technique. *Appl. Surf. Sci.* **238**, 184-188 (2004).

36. Xie, Y., Zheng, X., Jiang, X., Lu, J. & Zhu, L. Sonochemical synthesis and mechanistic
20

study of copper selenides $Cu_{2-x}Se$, β-CuSe, and $Cu_3Se_2$. *Inorg. Chem.* **41**, 387-392 (2002).

37. Yang, D., Zhu, Q., Chen, C., Liu, H., Liu, Z., Zhao, Z., Zhang, X., Liu, S. & Han, B. Selective electroreduction of carbon dioxide to methanol on copper selenide nanocatalysts. *Nat. Commun.* **10**, 1-9 (2019).

38. Zhang, Y., Zhang, C., Guo, Y., Liu, D., Yu, Y. & Zhang, B. Selenium vacancy-rich $CoSe_2$ ultrathin nanomeshes with abundant active sites for electrocatalytic oxygen evolution. *J. Mater. Chem. A* **7**, 2536-2540 (2019).

39. Sun, Y., Wu, X., Yuan, L., Wang, M., Han, M., Luo, L., Zheng, B., Huang, K. & Feng, S. Insight into the enhanced photoelectrocatalytic activity in reduced $LaFeO_3$ films. *Chem. Commun.* **53**, 2499 (2017).

40. Chakraborty, B., Beltrán-Suito, R., Hlukhyy, V., Schmidt, J., Menezes, P. W. & Driess, M. Crystalline copper selenide as a reliable non-noble electro(pre)catalyst for overall water splitting. *ChemSusChem* **13**, 3222-3229 (2020).

41. Masud, J., Liyanage, W. P., Cao, X., Saxena, A. & Nath, M. Copper selenides as high-efficiency electrocatalysts for oxygen evolution reaction. *ACS Appl. Energy Mater.* **1**, 4075-4083 (2018).

42. Shitu, I. G., Liew, J. Y. C., Talib, Z. A., Baqiah, H., Awang Kechik, M. M., Ahmad Kamarudin, M., Osman, N. H., Low, Y. J. & Lakin, I. I. Influence of irradiation time on the structural and optical characteristics of CuSe nanoparticles synthesized via microwave-assisted technique. *ACS omega* **6**, 10698-10708 (2021).